\documentclass[dvips,10pt]{article}

\usepackage{pstricks}
\usepackage{amssymb}
\usepackage{amstext}
\usepackage{amsmath}
\usepackage{txfonts}

\usepackage{xy}
\xyoption{all}

 \newcommand{\cuu}[1]{\Lbag{#1}\Rbag}

\newcommand{\truss}[2]{\underset{#2}{\stackrel{#1}{\rightarrow}}}
\newcommand{\trusss}[2]{\xrightarrow[#2]{#1}}

\newcommand{\Obj}{{\OOO}}
\newcommand{\Sbj}{{\SSS}}
\newcommand{\Ent}{{\sf E}}
\newcommand{\Rat}{{\sf R}}
\newcommand{\Bin}{{\sf B}}

\renewcommand{\to}{\rightarrow}
\newcommand{\ot}{\leftarrow}

\newcommand{\epi}{\twoheadrightarrow}
\newcommand{\ipe}{\twoheadleftarrow}

\newcommand{\id}{{\rm id}}

\newcommand{\AAA}{{\cal A}}
\newcommand{\BBB}{{\cal B}}

\newcommand{\OOO}{{\cal O}}

\newcommand{\SSS}{{\cal S}}

\renewcommand{\Bbb}{\mathbb}

\newcommand{\MMm}{{\Bbb M}}

\newcommand{\RRr}{{\Bbb R}}

\mathcode`\<="4268 %left delimiter
\mathcode`\>="5269 %right delimiter
\mathchardef\gt="313E %relation >
\mathchardef\lt="313C %relation <

 \def\pushright#1{{%              set up
    \parfillskip=0pt            % so \par doesnt push \square to left
    \widowpenalty=10000         % so we dont break the page before \square
    \displaywidowpenalty=10000  % ditto
    \finalhyphendemerits=0      % TeXbook exercise 14.32
    \leavevmode                 % \nobreak means lines not pages
    \unskip                     % remove previous space or glue
    \nobreak                    % don't break lines
    \hfil                       % ragged right if we spill over
    \penalty50                  % discouragement to do so
    \hskip.2em                  % ensure some space
    \null                       % anchor following \hfill
    \hfill                      % push \square to right
    {#1}                        % the end-of-proof mark (or whatever)
    \par}}                      % build paragraph

 \def\qed{\pushright{$\square$}\penalty-700 \smallskip}

\newenvironment{prf}[1]{\begin{trivlist} \item[{\bf ~Proof}#1.]}%
{\qed\end{trivlist}}

\newcommand{\be}[1]{\begin{#1}}

\newcommand{\ee}[1]{\end{#1}}
\newcommand{\beq}{\begin{equation}}
\newcommand{\eeq}{\end{equation}}
\newcommand{\ba}[1]{\begin{array}{#1}}
\newcommand{\ea}{\end{array}}
\newcommand{\bea}{\begin{eqnarray}}
\newcommand{\eea}{\end{eqnarray}}
\newcommand{\bear}{\begin{eqnarray*}}
\newcommand{\eear}{\end{eqnarray*}}
\newcommand{\bpr}{\begin{prf}{}}
\newcommand{\epr}{\end{prf}}
\newcommand{\bprf}[1]{\begin{prf}{#1}}
\newcommand{\eprf}{\end{prf}}

\newtheorem{thm}{Theorem}%[section]
\newtheorem{lemma}[thm]{Lemma}
\newtheorem{corollary}[thm]{Corollary}

\newtheorem{cond}{}[thm]

\newtheorem{prenumb}[thm]{\hspace{-1ex}}

\renewcommand{\paragraph}[1]{\smallskip\noindent{\bf #1} }

\title{Quantifying and qualifying trust:\\ Spectral decomposition of trust networks}

\author{Dusko Pavlovic\\%
Universities of Oxford and Twente\\
\small Email:~dusko\char64{comab.ox.ac.uk}
}

\date{}

\begin{document}
\maketitle

\begin{abstract}
In a previous FAST paper, I presented a quantitative model of the process of trust building, and showed that trust is accumulated like wealth: the rich get richer. This explained the pervasive phenomenon of adverse selection of trust certificates, as well as the fragility of trust networks in general. But a simple explanation does not always suggest a simple solution. It turns out that it is impossible to alter the fragile distribution of trust without sacrificing some of its fundamental functions. A solution for the vulnerability of trust must thus be sought elsewhere, without tampering with its distribution.
This observation was the starting point of the present paper. It explores a different method for securing trust: not by redistributing it, but by mining for its sources. The method used to break privacy is thus also used to secure trust. A high level view of the mining methods that connect the two is provided in terms of \emph{similarity networks}, and \emph{spectral decomposition} of  similarity preserving maps. This view may be of independent interest, as it uncovers a common conceptual and structural foundation of mathematical classification theory on one hand, and of the spectral methods of graph clustering and data mining on the other hand.
\end{abstract}

\section{Introduction}\label{Introduction}
\subsection{What is trust?}\label{What}
\textbf{Trust is an internal assumption of honesty.} In protocol analysis, we often assume that a principal Bob is {\em honest}. This usually means that Bob acts according to the prescriptions of a role in a given protocol. The notion of trust internalizes this assumption as a belief of another protocol participant Alice. We say that {\em Alice trusts Bob}\/ when she believes that Bob will act according to a given protocol; e.g., 
\begin{itemize}
\item when Alice is a shopper, she trusts that the shop Bob will deliver the goods;
\item when Alice is a shop, she trusts that the shopper Bob will pay for the goods;
\item when Alice uses public key infrastructure, she trusts that Bob's key is not compromised;
\item in an access control system, Alice trusts that Bob will not abuse resources.
\end{itemize}

\subsubsection{Trust process}
\begin{figure}[htbp]
\begin{center}
\def\JPicScale{.75}
\input{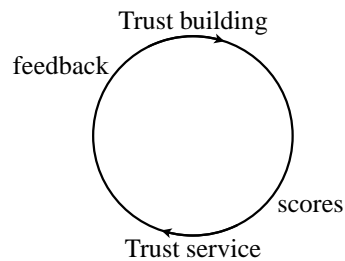}
\caption{Trust cycle}
\label{trust-cycle}
\end{center}
\end{figure}

In economics and game theory, trust is the instrument for transitioning from static rationality of non-cooperative behaviors to dynamics of cooperation \cite{BergJ:reciprocity}. To limit the risk exposure of the participants, trust is built gradually, keeping the cost of an attack at each point below a certain threshold of risk. Once the accumulated trust is above the required threshold, then it can be used as a basis for cooperation. Trust can thus be viewed as a process in two phases, {\em trust building\/} and {\em trust service}, that alternate as on Fig.~\ref{trust-cycle}.
\newcommand{\protocol}{\mbox{Trust building}}
\newcommand{\attack}{\mbox{Trust service}}
\newcommand{\downn}{scores}
\newcommand{\upp}{feedback}
%\begin{block}{Evolutionary cycle}
\vspace{-1\baselineskip}
\begin{itemize}
\item\emph{Trust building} is a process of {\em incremental\/} testing of another party's honesty, {\em viz\/} her readiness to cooperate within the framework of a given protocol. The output of this process are the trust scores, which record whether the party behaved honestly in the past transactions. This process was analyzed in \cite{PavlovicD:FAST08}. 

\item \emph{Trust service} uses the trust scores to guide some further transactions: a higher trust score attracts more transactions. The feedback about their outcome can be used to update the trust scores. 
\end{itemize}

\noindent{\it Remarks.}
There is a sense in which the process of trust building can be viewed as a foundation for authentication. When realized by means of a protocol, authentication is always based on a secret ("something you know"), or a token ("something you have"), or a biometric property ("something you are"). But the secret, the token, or the property must be previously authenticated to assure the current authentication. Every authentication must be preceded by another authentication. So a formal authentication is then a process of infinite regression,  "turtles all the way down"\footnote{The idea of a world standing on the back of a giant turtle is attributed to early Hindu philosophers. The idea that the turtle may be supported by another turtle, and so on, "turtles all the way down", is attributed to a mythical old lady who attended a lecture of a mythical philosopher, varying from William James to Bertrand Russell. This idea may have been sparked by Lewis Carroll's use of tortoise in the regression arguments \cite{CarrollL:tortoise}.}?! The notion of trust resolves this infinite regression. The process of trust building can be viewed as a primordial form of authentication, usually very weak, traversed in small, risk limiting steps. Of course, it generates trust, rather than a secret; but a first authentication protocol between two parties, which does generate a secret, is usually based on some form of trust; and the further, stronger secrets are then derived from each other. Indeed, the first SSH tunnel to an unknown server requires a leap of trust from the client; the first contact with a new certificate authority can only be based on trust, and so on.

Trust service has many different forms, which lead to the diverse forms of trust research. In the simplest case, a user records her own experience in an \emph{individual} trust vector, and uses it to make her own choices. In large networks, and in processes with many participants, a wider range of choices is made possible by a trust service which records \emph{collaborative} trust vectors, and supplies trust guidance for users and providers that had no previous experience with each other. This leads to the distinction between direct and indirect trust  \cite{BBK,YKB,JosangA:transitive}. In practice, indirect trust can be extracted from user feedback submitted to a trust server; or by surveillance and interpretation of the network structure and transactions: e.g., a return customer conveys trust by returning to the same shop, whereas she conveys dissatisfaction by returning the purchased goods.  Each of the approaches to aggregating indirect trust has its vulnerabilities and shortcomings. On the other hand, indirect trust service is indispensable not only for web commerce, but also for Public Key Infrastructure \cite{JosangA:PGP,ReiterM:Metric}.

\subsection{The problem of trust and the paradox of trust services}\label{Paradox}
Breaches of trust have been one of the hardest problems of social interaction since the dawn of mankind. Treason and betrayal attract the harshest punishments in all civilizations; and Dante reserved for these sins the deepest, Ninth Circle of Hell. But the problem of trust is not that it can be breached and abused. That is a part of its normal functioning. The problem of trust is that it can be farmed, transferred, marketed, and hijacked for the purpose of being breached and abused.

With the advent of networks and web commerce, the social processes of trust extend into cyberspace, engineered into various \emph{trust services}, including feedback, recommender, and reputation systems \cite{GuhaR:propagation,BoydC:trust-survey,Garcia-Molina:eigentrust,VarianH:recommender}, as well as the public infrastructures and web of trust \cite{JosangA:PGP}. This escalates the problem of trust to a new level. Oversimplifying a little, the problem can now be stated as a "paradox", namely that
\begin{itemize}
\item trust is not transferrable, but that
\item trust services must transfer trust.
\end{itemize}

On one hand, Alice's trust that Bob is honest is distilled from the experience of their past interactions, in order to guide their future interactions.  If the purpose of trust is thus to record Alice's view of Bob's behavior, then transferring trust defeats this purpose. Indeed, if Bob has been honest towards Alice, it does not mean that his friend Dave will also be honest, or that Bob will also be honest towards Alice's friend Carol; \emph{or} that Bob will be honest in a different kind of interaction.

On the other hand, in a large network, only a small portion of the participants can be expected to have direct previous interactions.  For all others, the trust guidance must be somehow extrapolated. This is the task of a trust service. So if Alice and Bob never met,  the trust service must derive a prediction of a future trust rating between them from Carol and Dave's available trust data, assuming that these two met, and that they are somehow related to Alice and Bob. So a trust service must transfer trust.

The compromises needed to resolve this paradox of trust lead to insecure trust services. But the vulnerabilities of trust services are different from the vulnerabilities usually studied in security research, where distinct attackers launch structured attacks. We first describe some examples of transferrable trust, and then their vulnerabilities.

\subsubsection{The simplest forms of trust transfer} 
The most general form of transferrable trust is \emph{reputation}. Bob's reputation is a statement about his honesty established by freely sharing the informations from the participants who interacted with him. If Alice never met Bob, she can rely upon his reputation, built and distributed in a peer to peer process, or by a devoted reputation service. E.g., Google's PageRank algorithm  \cite{brin98anatomy} can be viewed as a reputation service, derived by interpreting the hyperlinks as implicit trust statements \cite{Langville06google-book,PavlovicD:CSR08}. The main idea of Google's trustworthy search is to present the pages where the queried keyword is found ordered according to their reputation.

Another simple method of trust transfer are \emph{trust certificates}. Instead of keeping her trust in Bob's honesty for herself, Carol states this in a trust certificate, which she gives to Bob. Bob shows the certificate to Alice, and if Alice trusts Carol's recommendations, she will also trust Bob.

%\vspace{-\baselineskip}
\subsubsection{Adverse selection} 
Adverse selection is perhaps the most striking manifestation of the problem of trust. In the social processes of reputation, it manifests itself through the observations that "the pillars of the society" (to use the title of Ibsen's play \cite{IbsenH:pillars}), the most trusted social hubs, seem more likely to turn out malicious or corrupt than the average members of the group. In online trust services, this moral problem from old novels and stage tragedies becomes a technical problem for security researchers and engineers.  It also becomes a harder and more pernicious problem, because the commercial aspects of trust services provide additional incentives for strategic behavior, and concrete payoffs for focused attacks.

The adverse selection of trust certificates in web commerce was empirically documented in \cite{EdelmanB:adverse}. Through a battery of measurements and statistical analyses, it has been established that the web merchants with trust certificates are on the average twice as likely to scam their customers as those without such certificates. The claim was confirmed, with inessential variations, for all major trust certificate issuers. The phenomenon does not seem to be due to a lack of diligence, conflict of interest, or to a conspiracy between the merchants and the trust authorities, as it persists for other forms of trust certification. E.g., the sponsored links served by any of major search engines in response to a query are also on the average twice as likely to lead to scam as the organic results served in response to the same query, although the former are implicitly certified by the engine.

An explanation of adverse selection in terms of a basic model of dynamics of trust building was offered in \cite{PavlovicD:FAST08}. Since trust is built by updating the trust ratings following the positive or the negative interactions with the selected servers, and the selection of these servers is based on the previous trust ratings, it follows that trust attracts more trust: the servers with a higher trust rating are more likely to attract transactions, and thus to subsequently accumulate more trust, provided that they remain honest. This "rich get richer" schema \cite{Pareto} results in a {\em power law}, or {\em scale-free\/} distribution of trust ratings \cite{mitzenmacher04history,newman05zipf}. Its structural consequence is that there is a significant "heavy tail", consisting of the servers with very high trust ratings. Such distributions are well known to be robust under random perturbations, but extremely vulnerable to adaptive attacks \cite{DoyleJ:robust}. Intuitively, theft is more attractive and more harmful if very wealthy victims are available.

The fragility of scale-free networks, and the vulnerabilities arising from their distribution, cannot be mitigated by direct policy measures that would change the network structure. The power law distributions are not just a source of vulnerabilities, but also the basis of robustness of some networks. Such distributions are a pervasive manifestation of evolutionary dynamics of networks that arise from biological, social and economic processes. Modifying this dynamics is not a way to security. The heavy tails of wealth were, of course, redistributed many times throughout history, but they always re-emerged. The problems of heavy tails and adverse selection, and the problem of trust, require a different solution.

%The accumulation processes that lead to the power law distributions of wealth, trust, evolutionary fitness, word frequency, etc., are regulated by many subtle feedback loops\footnote{E.g., when some trusted providers turn out to be untrustworthy, and the trust service becomes unreliable, the process of trust building is slowed down and subdivided into smaller, less risky steps.}. We have no idea how to retune these feedback loops. 
%
%
%

\subsection{Mining for trust}
The data mining methods are often viewed as the main threat to privacy. We explore ways to make trust less abstract, and thus more secure, using the same methods. The idea is that trust and privacy are, in a sense, two sides of the same coin: the need for privacy arises from a lack of trust. Following this crude idea, we mine trust concepts from trust scores, to better control the trust transfer and to fine tune the trust services.

\subsubsection*{Related work} 
Initially, the trust relationships were analyzed mainly in the frameworks of access control \cite{Benantar:AC,LampsonB:AC} and public key infrastructure  \cite{BBK,YKB,LevienR,MaurerU:Trust,ReiterM:Metric}. 
The idea of trust network, and the elements of a probabilistic analysis of trust dynamics, go back to \cite{BBK}. The emergence of peer-to-peer and business-to-business service networks reawoke interest for this line of research, and amplified its impact \cite{GuhaR:propagation,Garcia-Molina:eigentrust,Karabulut,Garcia-Molina:Taxonomy}.

A different family of methods has been used to analyze the logical structure of the trust relationships, which are now viewed as the statements of principals' beliefs about each other. The various forms of uncertainty arising in such beliefs lead to the various nonstandard logical features \cite{SassoneV:Trust,GuttmanJ:Trust,JosangA:Subjlog,LampsonB:AC,Ninghui:Trust,Ninghui:JACM,EtalleS:trust}. 

The two types of trust research can be viewed as two projections of the trust network structure. The former type studies {\em global\/} dynamics of the trust relations in the form $a\truss {} r \ell$, projecting away\footnote{In \cite{LevienR,MaurerU:Trust,ReiterM:Metric} the trust concepts are present, but they carry very little structure: they are just labels used to distinguish, e.g., the delegation certificates from the binding certificates.} the trust concepts $\Phi$. The latter type studies logics of trust and the {\em local\/} trust relations in the form $a\truss \Phi {} \ell$ as a rely-guarantee exchange, projecting away the trust ratings $r$. Our previous analysis in \cite{PavlovicD:FAST08} was of the former type, as we ignored trust concepts and focused on quantitative dynamics of trust ratings. In the present paper we show how the qualitative distinctions of trust concepts $\Phi$ naturally arise from this quantitative dynamics of trust ratings $r$. The idea to mine them using the Singular Value Decomposition \cite[Sec.~5.4.5]{Golub-vanLoan} can be viewed as an extension of Latent Semantic Analysis \cite{LSI} to the domain of trust.

\subsubsection*{Outline of the paper} 
In Sec.~\ref{Graphs}, we motivate by a toy example and introduce the formal framework of trust graphs. In Sec.\ref{clusters} we spell out the connection between the trust formalism and a similarity formalism, which explains and justifies the methods we use to mine trust concepts from trust data. The main formal result is Prop.~\ref{sim-decomp}. Its main consequences are drawn in Sec.~\ref{qualified}. The results are applied on the toy example in Sec.\ref{Toy}. As always, the final section contains a summary of the result and some comments about the future work.

\section{Trust graphs}\label{Graphs}
In web commerce and social networks, trust services are provided through feedback, recommender, and reputation systems \cite{GuhaR:propagation,BoydC:trust-survey,Garcia-Molina:eigentrust,VarianH:recommender}, deployed as an integral part of a wider environment (such as Netflix, Amazon, eBay, or Facebook), or as a devoted service (on the span from TrustE and Verisign, to Epinions and Yelp). They are the cyberspace complement of the social processes of trust. The central component of most such systems is a network of users and providers, connected by trust ratings. This is what we call a \emph{trust network}. Its purpose is to support transfer of trust, and supply trust recommendations, or support or refine the existing trust scores. 

\subsection{A toy example}
\newcommand{\one}{\scriptscriptstyle 1.25}
\newcommand{\two}{\scriptscriptstyle .83}
\newcommand{\three}{\scriptscriptstyle 1.05}
\newcommand{\four}{\scriptscriptstyle 1.13}
\newcommand{\five}{\scriptscriptstyle .35 }
\newcommand{\six}{\scriptscriptstyle 1.12}
\newcommand{\seven}{\scriptscriptstyle 1.02}
\newcommand{\eight}{\scriptscriptstyle .21}
\newcommand{\nine}{\scriptscriptstyle 1.57}
\newcommand{\ten}{\scriptscriptstyle .35 }
\newcommand{\eleven}{\scriptscriptstyle -.56 }
\newcommand{\twelve}{\scriptscriptstyle .18}
\newcommand{\thirteen}{\scriptscriptstyle 1.02}
\newcommand{\fourteen}{\scriptscriptstyle .98 }
\newcommand{\fifteen}{\scriptscriptstyle -.12 }
\newcommand{\sixteen}{\scriptscriptstyle 0 }
\newcommand{\trustees}{$\Obj$}
\newcommand{\trustors}{$\Sbj$}
\newcommand{\ha}{a}
\newcommand{\hb}{b}
\newcommand{\hc}{c}
\newcommand{\hd}{d}
\newcommand{\he}{e}
\newcommand{\hI}{{i}}
\newcommand{\hII}{{j}}
\newcommand{\hIII}{{k}}
\newcommand{\hIV}{{\ell}}
An example of  a trust network is given in Fig.~\ref{net}. 
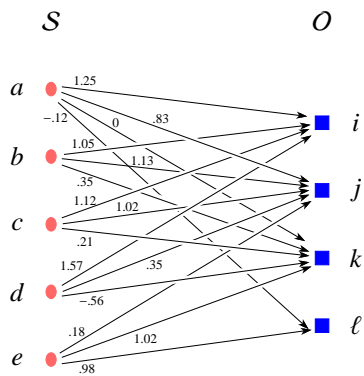
\begin{figure}[htbp]
\begin{center}
\def\JPicScale{.45}
\ifx\JPicScale\undefined\def\JPicScale{1}\fi
\psset{unit=\JPicScale mm}
\psset{linewidth=0.3,dotsep=1,hatchwidth=0.3,hatchsep=1.5,shadowsize=1,dimen=middle}
\psset{dotsize=0.7 2.5,dotscale=1 1,fillcolor=black}
\psset{arrowsize=1 2,arrowlength=1,arrowinset=0.25,tbarsize=0.7 5,bracketlength=0.15,rbracketlength=0.15}
\begin{pspicture}(0,0)(110,105)
\psline[border=0.75,arrowscale=1 1.45,arrowsize=1.45 2,arrowlength=1.55]{->}(22,81)(96,14)
\pspolygon[linecolor=blue,fillcolor=blue,fillstyle=solid](98,53)(102,53)(102,57)(98,57)
\newrgbcolor{userLineColour}{1 0.4 0.4}
\newrgbcolor{userFillColour}{1 0.4 0.4}
\rput{90}(20,45){\psellipse[linecolor=userLineColour,fillcolor=userFillColour,fillstyle=solid](0,0)(2,1.5)}
\pspolygon[linecolor=blue,fillcolor=blue,fillstyle=solid](98,33)(102,33)(102,37)(98,37)
\pspolygon[linecolor=blue,fillcolor=blue,fillstyle=solid](98,73)(102,73)(102,77)(98,77)
\newrgbcolor{userLineColour}{1 0.4 0.4}
\newrgbcolor{userFillColour}{1 0.4 0.4}
\rput{90}(20,65){\psellipse[linecolor=userLineColour,fillcolor=userFillColour,fillstyle=solid](0,0)(2,1.5)}
\psline[border=0.75,arrowscale=1 1.45,arrowsize=1.45 2,arrowlength=1.55]{->}(23.12,85.62)(95.62,76.88)
\psline[arrowscale=1 1.45,arrowsize=1.45 2,arrowlength=1.55]{->}(23.12,64.38)(96.25,56.25)
\psline[border=1.1,arrowscale=1 1.45,arrowsize=1.45 2,arrowlength=1.55]{->}(23.12,62.5)(96.25,36.88)
\rput[t](30,59){$\five$}
\newrgbcolor{userLineColour}{1 0.4 0.4}
\newrgbcolor{userFillColour}{1 0.4 0.4}
\rput{90}(20,85){\psellipse[linecolor=userLineColour,fillcolor=userFillColour,fillstyle=solid](0,0)(2,1.5)}
\newrgbcolor{userLineColour}{1 0.4 0.4}
\newrgbcolor{userFillColour}{1 0.4 0.4}
\rput{90}(20,25){\psellipse[linecolor=userLineColour,fillcolor=userFillColour,fillstyle=solid](0,0)(2,1.5)}
\psline[border=0.75,arrowscale=1 1.45,arrowsize=1.45 2,arrowlength=1.55]{->}(23,82)(95,40)
\psline[border=0.75,arrowscale=1 1.45,arrowsize=1.45 2,arrowlength=1.55]{->}(22.5,6.88)(96.88,51.88)
\psline[border=0.75,arrowscale=1 1.45,arrowsize=1.45 2,arrowlength=1.55]{->}(23.12,46.88)(96.25,73.12)
\psline[border=0.75,arrowscale=1 1.45,arrowsize=1.45 2,arrowlength=1.55]{->}(23.38,45.12)(96.25,55)
\psline[border=0.85,arrowscale=1 1.45,arrowsize=1.45 2,arrowlength=1.55]{->}(23.75,65.62)(96.62,74.38)
\psline[border=0.75,arrowscale=1 1.45,arrowsize=1.45 2,arrowlength=1.55]{->}(23.12,23.75)(96.25,33.75)
\psline[border=0.75,arrowscale=1 1.45,arrowsize=1.45 2,arrowlength=1.55]{->}(23.12,26.88)(96.88,71.25)
\psline[border=0.75,arrowscale=1 1.45,arrowsize=1.45 2,arrowlength=1.55]{->}(23.38,25.12)(96.25,53.75)
\rput[b](30,50.62){$\six$}
\rput[b](42.5,48.75){$\seven$}
\rput[t](30,41.25){$\eight$}
\rput[br](29.38,31.25){$\nine$}
\rput[t](31.88,23.75){$\eleven$}
\rput[tl](48.12,34.38){$\ten$}
\rput(10,85){$\ha$}
\rput(10,65){$\hb$}
\rput(10,45){$\hc$}
\rput(10,25){$\hd$}
\rput(110,75){$\hI$}
\rput(110,55){$\hII$}
\rput(110,35){$\hIII$}
\rput[b](30,86){$\one$}
\rput[b](29.38,67.5){$\three$}
\pspolygon[linecolor=blue,fillcolor=blue,fillstyle=solid](98,13)(102,13)(102,17)(98,17)
\newrgbcolor{userLineColour}{1 0.4 0.4}
\newrgbcolor{userFillColour}{1 0.4 0.4}
\rput{90}(20,5){\psellipse[linecolor=userLineColour,fillcolor=userFillColour,fillstyle=solid](0,0)(2,1.5)}
\psline[border=0.75,arrowscale=1 1.45,arrowsize=1.45 2,arrowlength=1.55]{->}(23.12,3.75)(96.88,13.75)
\rput(10,5){$\he$}
\rput(110,15){$\hIV$}
\rput(100,105){\trustees}
\rput(20,105){\trustors}
\psline[border=0.75,arrowscale=1 1.45,arrowsize=1.45 2,arrowlength=1.55]{->}(23,84)(96.25,57.5)
\psline[border=0.75,arrowscale=1 1.45,arrowsize=1.45 2,arrowlength=1.55]{->}(23.12,43.75)(96.26,35.63)
\psline[border=0.75,arrowscale=1 1.45,arrowsize=1.45 2,arrowlength=1.55]{->}(23.12,5.62)(96.88,32.5)
\rput[br](30,11.88){$\twelve$}
\rput[t](30.62,3.75){$\fourteen$}
\rput[tl](44.38,12.5){$\thirteen$}
\rput[tr](25,78){$\fifteen$}
\rput[bl](50,75){$\two$}
\rput[Bl](38.12,73.75){$\sixteen$}
\rput[b](46.88,62.5){$\four$}
\end{pspicture}
\caption{A trust network}
\label{net}
\end{center}
\end{figure}
Let us suppose that it is derived from the market transactions in a small town, deep in the World Wild West. The market consists of four web shops, say $\Obj = \{i,j,k,\ell\}$, and five shopping agents $\Sbj = \{a,b,c,d,e\}$. A local bank clears all payments in town, and the diligent banker maintains the matrix of trust scores, like in Table~1.
\begin{table}[htdp]
\begin{center}
\begin{tabular}{|c||c|c|c|c|c|}
\hline
& a & b & c & d& e\\
\hline \hline
\ \ \ $i$\ \ \ \ & 1.25 & 1.05 & 1.12 & 1.57& \\
\hline
$j$ & .83 & 1.13 & 1.02 & .35 & .18 \\
\hline
$k$ & 0 & .35 & .21 & -.56 &  1.02\\
\hline
$\ell$ & -.12 &  & & &.98\\
\hline
\end{tabular}
\label{scores}
\end{center}
\caption{Trust scores}
\end{table}%
%\begin{table}[htdp]
%\begin{center}
%\begin{tabular}{|c||c|c|c|c|}
%\hline
%& a & b & c & d\\
%\hline \hline
%I & 1.25 & 1.05 & 1.12 & 1.57\\
%\hline
%II & .83 & 1.13 & 1.02 & .35 \\
%\hline
%k & 0 & .35 & .21 & -.56\\
%\hline
%\end{tabular}
%\end{center}
%\label{scores}
%\caption{A trust matrix}
%\end{table}%
Suppose that the scores are derived from the total value of the transactions and from user feedback, in a uniform way. The fields are empty where there are no data. The negative entries result from negative feedback, or returned goods. The scale of the scores is irrelevant, provided they are all obtained in the same way.
The banker provides some trust recommendations, in support of a stable market. Some of the typical tasks are:
\begin{itemize}
\item \emph{Predict the missing trust scores:} E.g., how will the shopper $d$ like the shop $\ell$?  
\item \emph{Refine the existing trust scores:} Suppose that the merchant $i$ acquires the shop $j$, with the intention to deliver the same services through both outlets. Which shop should now be recommended to the shopper $b$ when he decides to fill her virtual pantry?
\end{itemize}
We study the second type of query here in detail. The approach to the first query is sketched in the final section, but the details must be left for the sequel to this paper.
%
%
%We study a method to address the second query in some detail, and sketch an approach to the first query.
%

\subsection{Formalizing trust}\label{Formalizing}
Formally, and a little more generally, a trust score, or trust statement, can be viewed as a quadruple $a\truss \Phi r \ell$, where
\begin{itemize}
\item $a$ is the {\em trustor},
\item $\ell$ is the {\em trustee},
\item $\Phi$ is the {\em entrusted protocol (concept, property)}, and
\item $r$ is the {\em trust rating}.
\end{itemize}
The trust statements $a\truss \Phi r \ell$ can be viewed as the edges of a bipartite labelled graph, i.e. the elements of a set $\Bin$ given with the structure
\bear
\MMm   & = &  \big(\Ent\times \Rat \ot \Bin \to \Sbj\times\Obj \big)
\eear
where 
\begin{itemize}
\item $\Sbj$ and $\Obj$ are the sets of trustors and trustees (or subjects and objects) respectively,
\item $\Ent$ is a lattice of entrusted concepts,
\item $\Rat$ is an ordered ring of ratings, usually the field of real numbers.
\end{itemize}
We call such a bipartite graph a \emph{trust graph}. This is what a trust services works with.

\subsection{Idea: Qualifying trust}\label{Need}
In a trust statement $a\truss \Phi r \ell$, the trust score $r$ quantifies $a$'s trust for $\ell$. The trust concept $\Phi$ qualifies it. Qualified trust is less abstract, and thus less vulnerable to unwarranted transfers. We develop an algebra of trust concepts, as a tool for mitigating the vulnerabilities of trust transfer.

The strategy is to first mine for trust cliques from the trust graph data. Intuitively, a trust clique is a set of trustees who trust the same trustors; or a set of trustors trusted by the same trustees. In the second step, a \emph{trust concept} will be defined as a pair of trust cliques, one of trustors, one of trustees, correlated by trust. An abstract framework for clustering the trustors and the trustees sets the stage for recognizing the trust cliques and concepts.

\section{Clusters and concepts}\label{clusters}
\subsection{Similarity networks}
\be{defn}
A \emph{similarity network} is a set $A$ together with a \emph{similarity map} $s=s_A : A\times A\to [0,1]$
such that for all $x,y\in A$ holds
\[
s(x,x) =  1 \qquad \qquad s(x,y) = s(y,x)
\]
A \emph{similarity morphism} between similarity networks $A$ and $B$ is a map $f:A\to B$ such that for  all $x,y\in A$ holds
\bear
s_A(x,y) &\leq & s_B(fx,fy)
\eear
\end{defn} 

\be{defn}
For similarity networks $A$ and $C$, a \emph{clustering} $c:A\epi C$ is a pair of similarity morphisms $c = <c',c''>$, where
\begin{itemize}
\item $c':C\to A$ satisfies $s_C(x,y) = s_A(c'x,c'y)$ for all $x,y\in C$, and
\item $c'':A\to C$ satisfies $c''\circ c' = \id_C$.
\end{itemize}
\ee{defn}

Intuitively, the surjection $c'':A\to C$ maps each node $z\in A$ into a unique cluster $c''z\in C$, whereas the injection $c':C\to A$ picks a representative $c'x\in A$ from each cluster $z\in C$, in such a way that the similarity of the clusters $s_C(x,y)$ is just the similarity of their representatives $s_A(c'x, c'y)$. 

Some similarity morphisms determine a canonical clustering for both networks that they connect.

\be{defn}
The \emph{spectral decomposition} of a similarity morphism $f:A\to B$ is a pair of clusterings $A\stackrel u \epi \cuu f \stackrel v \ipe B$ such that 
\begin{gather*}
f \ = \  v'\circ u'' \\
\xymatrix@C3pc@R3pc{
A \ar[rr]^f \ar@/_/
@{->>}[dr]_{u''}  &&  B \ar@/_/
@{->>}[dl]_{v''} \\
 & \cuu f  \ar@/_/
 @{^{(}->}[ul]_{u'} \ar@/_/
 @{^{(}->}[ur]_{v'}
}\end{gather*}
The similarity network $\cuu f$ is the \emph{cluster range} of $f$.
\ee{defn}

\begin{lemma}\label{unique}
Whenever it exists, the spectral decomposition of is unique up to a similarity isomorphism. More precisely, if both $A\stackrel u \epi H \stackrel v \ipe B$ and $A\stackrel s \epi K \stackrel t \ipe$ are spectral decompositions of the same similarity morphism, then there is an isomorphism $\iota:H\to K$ such that $u' = s'\circ \iota$ and $v' = t' \circ \iota$ (which implies $s'' = \iota \circ u''$ and $t'' = \iota \circ v''$).
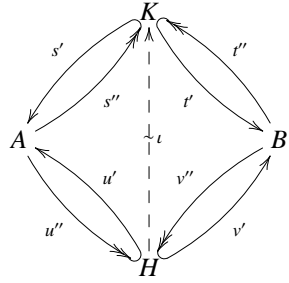
\[\xymatrix@C3pc@R3pc{
 & K  \ar@/_/
 @{^{(}->}[dl]_{s'} \ar@/_/
 @{^{(}->}[dr]_{t'} \\
A %\ar[rr]^f 
\ar@/_/
@{->>}[dr]_{u''} \ar@/_/@{->>}[ur]_{s''} &&  B \ar@/_/
@{->>}[dl]_{v''} \ar@/_/@{->>}[ul]_{t''} \\
 & H\ar@{-->}[uu]|{\sim}_{\iota}  \ar@/_/
 @{^{(}->}[ul]_{u'} \ar@/_/
 @{^{(}->}[ur]_{v'}
}\]
\end{lemma}

\subsection{From trust graph to similarity networks}\label{from}
The spectral decomposition will turn out to be useful for mining and extrapolating trust concepts in a trust graph. At this stage, though, we ignore the task of extrapolating the missing trust scores, and restrict attention to complete submatrices of the given trust graph. E.g., from Table~1, we first just take the matrix
\bear
M & = & \begin{pmatrix}
1.25 & 1.05 & 1.12 & 1.57\\
.83 & 1.13 & 1.02 & .35 \\
0 & .35 & .21 & -.56
\end{pmatrix}
\eear
corresponding to the subnetwork spanned by the trustors $a,b,c,d$ and the trustees $i,j,k$. Over these two sets, we first form vector spaces $\RRr^{\{a,b,c,d\}}$ and $\RRr^{\{i,j,k\}}$. We are actually only interested in unit vectors, i.e. in the sets 
\bear
\Sigma & = & \{\varphi \in \RRr^{\{a,b,c,d\}}\ |\ |\varphi | = 1\}\\
\Theta & = & \{\vartheta \in \RRr^{\{i,j,k\}}\ |\ |\vartheta | = 1\}
\eear
The elements of $\Sigma$ represent the communities of trustors; the elements of $\Theta$ are the communities of trustees. The components of a community $\varphi = \begin{pmatrix}\varphi_a &  \varphi_b & \varphi_c & \varphi_d\end{pmatrix}^\top$ quantify the participation of each member. Both $\Sigma$ and $\Theta$ can be naturally viewed as similarity networks, as the similarity of two communities can be measured using their inner product:
\[
s_\Sigma(\varphi, \psi) = |<\varphi|\psi>|\qquad \qquad 
s_\Theta(\vartheta, \tau) = |<\vartheta|\tau>|
\]
\paragraph{Remark.} The absolute value in the definition of similarity means that, strictly speaking, each community is represented by two unit vectors $\varphi$ and $-\varphi$, which are indistinguishable since $s_\Sigma(\varphi, \psi) = s_\Sigma(-\varphi, \psi)$ for all $\psi\in \Sigma$. Geometrically, a community is thus actually not a unit vector, but a ray, i.e. the 1-dimensional subspace generated by it. \emph{When the confusion is unlikely, we tacitly identify rays and the pairs of unit vectors that represent them.}

All this begins to make sense when we observe that the linear operator $M:\RRr^{\{a,b,c,d\}} \to \RRr^{\{i,j,j\}}$, determined by the matrix  $M$ of trust scores, induces a map $\widehat  M:\Sigma \to \Theta$, defined
\bear
\widehat M\varphi & = & \frac{M\varphi}{|M\varphi|}
\eear
But unfortunately, this is not a similarity morphism. E.g., it strictly decreases the similarity of 
%$\varphi =  \smash \begin{pmatrix} \scriptstyle 0 \\ \scriptstyle .5 \\ \scriptstyle .3 \\ \scriptstyle -.8\end{pmatrix}$  and $\psi = \begin{pmatrix} \scriptstyle .25 \\ \scriptstyle .5 \\  \scriptstyle .4 \\  \scriptstyle -.15\end{pmatrix}$.
\[
\varphi = \begin{pmatrix} 0 \\ .5 \\ .3 \\ -.8\end{pmatrix} \qquad \mbox{ and } \qquad \psi = \begin{pmatrix} .25 \\ .5 \\ .4 \\ -.15\end{pmatrix}\] 

\subsection{Concepts generate clusters}
The following definition formalizes the intuition from Sec.~\ref{Need}.

\be{defn}
Let $\AAA$ and $\BBB$ be finite dimensional real vector spaces, $M:\AAA\to \BBB$ a linear operator, and $M^\top:\BBB\to \AAA$ its transpose. Let $A$ and $B$ be the sets of 1-dimensional subspaces, i.e. rays in $\AAA$ and $\BBB$ respectively. A pair $<\alpha, \beta>\in A\times B$ is called a \emph{concept} with respect to $M$ if $M\alpha = \beta$ and $M^\top \beta = \alpha$. The \emph{concept spectrum} $\Ent_M$ is the set of all concepts with respect to $M$. 
\end{defn}

%\be{defn}
%Let $\MMm$ be a complete trust graph, i.e. such that there is exactly one edge $a \truss {} r  \ell$ for every $a\in \Sbj$ and $\ell \in \Obj$. Let $F$ be the corresponding trust matrix.
%
%We call \emph{trust communities} the  rays, i.e. 1-dimensional subspaces $\sigma \subseteq \RRr^\Sbj$ and $\omega \subseteq \RRr^\Obj$.
%
%A \emph{trust concept} is  a pair $<\sigma, \omega>$ of trust communities such that $F\sigma = \omega$ and $F^\top \omega = \sigma$. The \emph{trust spectrum} of the trust graph $\MMm$ is the set of all of its trust concepts.
%\end{defn}

Now we circumvent the obstacle noted at the end of the preceding Section, and relate concepts with clusters.

\be{proposition}\label{sim-decomp}
Let $\AAA$ and $\BBB$ be finite dimensional vector spaces and $A$ and $B$ the induced similarity networks. Then every linear operator $M:\AAA\to \BBB$ induces a unique linear operator $F = F_M:\AAA\to \BBB$ such that
\begin{itemize}
\item the map $\widehat F : A \to B$, defined by $\widehat F \varphi = \frac{F\varphi}{|F\varphi|}$,  is a similarity morphism,
\item there is a spectral decomposition $A\stackrel u \epi \cuu{\widehat F} \stackrel v \ipe B$, and
\item the cluster range $\cuu{\widehat F}$ is generated by the concept spectrum $\Ent_M$, in the sense that every element of $\cuu{\widehat F}$ is  a convex combination of concepts with respect to $M$.
\end{itemize}
\end{proposition}

\bpr
By the Singular Value Decomposition \cite[Sec.~2.5.,5.4.5]{Golub-vanLoan}, the matrix $M$ can be decomposed in the form
\begin{gather*}
M  =  V \Lambda U^\top\\
\xymatrix@C3pc@R3pc{
\AAA \ar[rrrr]^M \ar@/_/
@{->>}[dr]_{U^\top}  && && \BBB \ar@/_/
@{->>}[dl]_{V^\top} \\
 & \hat{\AAA} \ar[rr]_{\Lambda} \ar@/_/
 @{^{(}->}[ul]_{U} & & \hat{\BBB} \ar@/_/
 @{^{(}->}[ur]_{V}
}\end{gather*}
where  \begin{itemize}
\item $U$ and $V$ are isometries, i.e. $U^\top U = \id$ and $V^\top V = \id$, whereas 
\item $\Lambda$ is a positive diagonal matrix. 
\end{itemize}
The latter implies that $\Lambda$ is an isomorphism, so we can take $\hat \AAA = \hat \BBB$ without loss of generality. The operator $F = F_M$ can now be defined
\begin{gather*}
F \ = \  V \circ U^\top \\
\xymatrix@C3pc@R3pc
{
\AAA \ar[rr]^F \ar@/_/
@{->>}[dr]_{U^\top}  &&  \BBB \ar@/_/
@{->>}[dl]_{V^\top} \\
 & \hat \AAA = \hat \BBB  \ar@/_/
 @{^{(}->}[ul]_{U} \ar@/_/
 @{^{(}->}[ur]_{V}
}\end{gather*}
The claims follow by passing from vector spaces to the induced similarity networks of rays. 
%The concepts are just the eigenspaces of $M$, i.e. the rays fixed by $\Lambda$.
\epr

\noindent\textbf{Comment.} The preceding Proposition has some interesting conceptual and technical repercussions, that may not be immediately obvious. Conceptually, it displays the connection between clustering in the style of mathematical taxonomy \cite{Jardine-Sibson}, and concept mining in the style of Latent Semantic Analysis \cite{LSI}. By showing how the Singular Value Decomposition extracts the similarity invariants, it explains why the abstract algebra of matrix decomposition yields semantically meaningful results. The proof also determines the formal sense in which the singular values are semantically irrelevant --- the statement often heard from the practitioners. We shall see below what kind of information the singular values do carry, and also how factoring them out opens up an alley towards gluing spectra, and towards extrapolating the missing trust scores. The technical repercussions of Prop.~\ref{sim-decomp} follow from there.

%Although trust matrices are always finite, it may be worth making a mental note that Prop.~\ref{sim-decomp} remains valid when $\AAA$ and $\BBB$ are separable Hilbert spaces, and $F$ a bounded map. The infinite dimensional approach may be worth looking at if the dimensions of the problem become unfeasibly large.
%
%

\subsection{Trust qualified}\label{qualified}
The first consequence of Prop.~\ref{sim-decomp} for trust is that any trust matrix $M$ can be decomposed into \emph{qualified} trust matrices. 

\begin{corollary}
The similarity preserving matrix $F$ induced by a matrix $M$ from Prop.~\ref{sim-decomp} decomposes in the form 
\bear
F & = & \sum_{\Phi\in \Ent_M} F_\Phi \mbox{ with} \\
F_\Phi & = & V_\Phi \circ U_\Phi^\top
\eear
where
\begin{itemize}
\item $\Ent_M$ is the concept spectrum of $M$, 
\item $V_\Phi$ and $U_\Phi$ denote the columns of $V$ and $U$.
\end{itemize}
The matrices $F_\Phi$ are called \emph{qualified trust matrices} induced by $M$. The matrix $M$ itself decomposes as
\bear
M & = & \sum_{\Phi\in \Ent_M} \lambda_\Phi F_\Phi
\eear 
where $\lambda_\Phi$ are its singular values.
\end{corollary}

\noindent\textbf{Comment.} Componentwise, the above decomposition means that each trust rating $M_{a\ell}$ is thus decomposed in the form
\bear
M_{a\ell} & = & \sum_{\Phi\in \Ent_M} r_\Phi\mbox{ with}\\
r_\Phi & = & \lambda_\Phi U_{a\Phi} V_{\ell\Phi}
\eear
The component $r_\Phi$ measures the contribution of the trust concept $\Phi$ to the trust rating $M_{a\ell}$. When $M_{a\ell}$ is viewed as the edge $a\truss{} r \ell$ of the corresponding trust graph, then the above decomposition becomes
\[
a \trusss{\sum_{\Phi} r_\Phi \Phi}{r = \sum_{\Phi} r_\Phi} \ell 
\]
The vector $\sum_{\Phi} r_\Phi \Phi$ is in the range space of $M$, generated by its concept spectrum.

\subsection{Mining for concepts in the World Wild West}\label{Toy}
To apply Prop.~\ref{sim-decomp}, the banker in our town in the World Wild West first decomposes the matrix $M$ from Sec.~\ref{from} to the form $M= V \Lambda U^\top$, where  
\[ V  =  
\begin{pmatrix}
.83 & -.4\\
.55 & .6  \\
0 & .7 
\end{pmatrix}
 \qquad
\Lambda  = 
\begin{pmatrix}
3 & 0\\
0 & 1
\end{pmatrix}\qquad
U  =  \begin{pmatrix}
.5 & 0 \\
.5 & .5 \\
.5 &.3\\
.5 & -.8 
\end{pmatrix}
\]
The trust graph corresponding to $M$ and its decomposition are displayed on Figures~\ref{fig-cgraph} and \ref{fig-decomp} (with the 0 links omitted).  
\begin{figure}[htdp]
\begin{minipage}[b]{0.4\linewidth}
\centering
\def\JPicScale{.45}
\ifx\JPicScale\undefined\def\JPicScale{1}\fi
\psset{unit=\JPicScale mm}
\psset{linewidth=0.3,dotsep=1,hatchwidth=0.3,hatchsep=1.5,shadowsize=1,dimen=middle}
\psset{dotsize=0.7 2.5,dotscale=1 1,fillcolor=black}
\psset{arrowsize=1 2,arrowlength=1,arrowinset=0.25,tbarsize=0.7 5,bracketlength=0.15,rbracketlength=0.15}
\begin{pspicture}(0,0)(110,90)
\pspolygon[linecolor=blue,fillcolor=blue,fillstyle=solid](98,38)(102,38)(102,42)(98,42)
\newrgbcolor{userLineColour}{1 0.4 0.4}
\newrgbcolor{userFillColour}{1 0.4 0.4}
\rput{90}(20,30){\psellipse[linecolor=userLineColour,fillcolor=userFillColour,fillstyle=solid](0,0)(2,1.5)}
\pspolygon[linecolor=blue,fillcolor=blue,fillstyle=solid](98,18)(102,18)(102,22)(98,22)
\pspolygon[linecolor=blue,fillcolor=blue,fillstyle=solid](98,58)(102,58)(102,62)(98,62)
\rput(100,90){\trustees}
\rput(20,90){\trustors}
\newrgbcolor{userLineColour}{1 0.4 0.4}
\newrgbcolor{userFillColour}{1 0.4 0.4}
\rput{90}(20,50){\psellipse[linecolor=userLineColour,fillcolor=userFillColour,fillstyle=solid](0,0)(2,1.5)}
\psline[border=0.75,arrowscale=1 1.45,arrowsize=1.45 2,arrowlength=1.55]{->}(23.12,70.62)(95.62,61.88)
\psline[arrowscale=1 1.45,arrowsize=1.45 2,arrowlength=1.55]{->}(23.12,49.38)(96,40)
\psline[border=1.1,arrowscale=1 1.45,arrowsize=1.45 2,arrowlength=1.55]{->}(23.12,47.5)(96.25,21.88)
\rput[t](30,44){$\five$}
\newrgbcolor{userLineColour}{1 0.4 0.4}
\newrgbcolor{userFillColour}{1 0.4 0.4}
\rput{90}(20,70){\psellipse[linecolor=userLineColour,fillcolor=userFillColour,fillstyle=solid](0,0)(2,1.5)}
\newrgbcolor{userLineColour}{1 0.4 0.4}
\newrgbcolor{userFillColour}{1 0.4 0.4}
\rput{90}(20,10){\psellipse[linecolor=userLineColour,fillcolor=userFillColour,fillstyle=solid](0,0)(2,1.5)}
\psline[border=0.75,arrowscale=1 1.45,arrowsize=1.45 2,arrowlength=1.55]{->}(23.12,69.38)(95.62,41.88)
\psline[border=0.75,arrowscale=1 1.45,arrowsize=1.45 2,arrowlength=1.55]{->}(23.12,28.12)(96.25,20)
\psline[border=0.75,arrowscale=1 1.45,arrowsize=1.45 2,arrowlength=1.55]{->}(23.12,31.88)(96.25,58.12)
\psline[border=0.75,arrowscale=1 1.45,arrowsize=1.45 2,arrowlength=1.55]{->}(23.38,30.12)(96.25,38.12)
\psline[border=0.85,arrowscale=1 1.45,arrowsize=1.45 2,arrowlength=1.55]{->}(23.75,50.62)(96.62,59.38)
\psline[border=0.75,arrowscale=1 1.45,arrowsize=1.45 2,arrowlength=1.55]{->}(23.12,8.12)(96.25,18.12)
\psline[border=0.75,arrowscale=1 1.45,arrowsize=1.45 2,arrowlength=1.55]{->}(23.12,11.88)(96.88,56.25)
\psline[border=0.75,arrowscale=1 1.45,arrowsize=1.45 2,arrowlength=1.55]{->}(23.38,10.12)(96.25,36.25)
\rput[b](30,35.62){$\six$}
\rput[b](43.75,33.12){$\seven$}
\rput[t](30,26.25){$\eight$}
\rput[b](29.38,16.88){$\nine$}
\rput[t](30,8){$\eleven$}
\rput[t](46.88,17.5){$\ten$}
\rput(10,70){$\ha$}
\rput(10,50){$\hb$}
\rput(10,30){$\hc$}
\rput(10,10){$\hd$}
\rput(110,60){$\hI$}
\rput(110,40){$\hII$}
\rput(110,20){$\hIII$}
\rput(57.5,85){}
\rput[b](44.38,61.88){$\two$}
\rput[b](30,71){$\one$}
\rput[b](30,53){$\three$}
\rput[b](44.38,47.5){$\four$}
\end{pspicture}
\caption{Trust graph $M$}
\label{fig-cgraph}
\end{minipage}
\hspace{0.2cm}
\renewcommand{\one}{\scriptscriptstyle .5}
\renewcommand{\two}{\scriptscriptstyle .5}
\renewcommand{\three}{\scriptscriptstyle .5}
\renewcommand{\four}{\scriptscriptstyle .5}
\renewcommand{\five}{\scriptscriptstyle .3}
\renewcommand{\six}{\scriptscriptstyle .5}
\renewcommand{\seven}{\scriptscriptstyle -.8}
\renewcommand{\eight}{\scriptscriptstyle .83}
\renewcommand{\nine}{\scriptscriptstyle .55}
\renewcommand{\ten}{\scriptscriptstyle -.4}
\renewcommand{\eleven}{\scriptscriptstyle .6}
\renewcommand{\twelve}{\scriptscriptstyle .7}
\newcommand{\One}{\scriptstyle 1}
\newcommand{\Three}{\scriptstyle 3}
\newcommand{\recommenders}{$\Ent_M$}
\newcommand{\concone}{\scriptstyle \Phi_1}
\newcommand{\conctwo}{\scriptstyle \Phi_2}
\begin{minipage}[b]{0.4\linewidth}
\centering
\def\JPicScale{.45}
\ifx\JPicScale\undefined\def\JPicScale{1}\fi
\psset{unit=\JPicScale mm}
\psset{linewidth=0.3,dotsep=1,hatchwidth=0.3,hatchsep=1.5,shadowsize=1,dimen=middle}
\psset{dotsize=0.7 2.5,dotscale=1 1,fillcolor=black}
\psset{arrowsize=1 2,arrowlength=1,arrowinset=0.25,tbarsize=0.7 5,bracketlength=0.15,rbracketlength=0.15}
\begin{pspicture}(0,0)(110,90)
\psline[border=0.75,arrowscale=1 1.45,arrowsize=1.45 2,arrowlength=1.55]{->}(65,51.88)(96.25,60)
\psline[arrowscale=1 1.45,arrowsize=1.45 2,arrowlength=1.55]{->}(23,69)(56,54)
\pspolygon[linecolor=blue,fillcolor=blue,fillstyle=solid](98,38)(102,38)(102,42)(98,42)
\pspolygon[linecolor=blue,fillcolor=blue,fillstyle=solid](98,18)(102,18)(102,22)(98,22)
\pspolygon[linecolor=blue,fillcolor=blue,fillstyle=solid](98,58)(102,58)(102,62)(98,62)
\rput(100,90){\trustees}
\rput(20,90){\trustors}
\rput(60,90){\recommenders}
\psline[border=0.75,arrowscale=1 1.45,arrowsize=1.45 2,arrowlength=1.55]{->}(65,30)(96.88,38.12)
\rput[b](27,52){$\two$}
\rput[t](27,28){$\five$}
\rput[b](28,68){$\one$}
\rput[b](27,18){$\six$}
\psline[border=0.75,arrowscale=1 1.45,arrowsize=1.45 2,arrowlength=1.55]{->}(64.38,32.5)(96.25,58.12)
\psline[border=0.75,arrowscale=1 1.45,arrowsize=1.45 2,arrowlength=1.55]{->}(65,48.12)(96.88,40.62)
\psline[border=0.75,arrowscale=1 1.45,arrowsize=1.45 2,arrowlength=1.55]{->}(64.38,27.5)(96.25,20)
\rput[b](90,59.38){$\eight$}
\rput(110,60){$\hI$}
\rput(110,40){$\hII$}
\rput(110,20){$\hIII$}
\newrgbcolor{userLineColour}{1 0.4 0.4}
\newrgbcolor{userFillColour}{1 0.4 0.4}
\rput{90}(20,30){\psellipse[linecolor=userLineColour,fillcolor=userFillColour,fillstyle=solid](0,0)(2,1.5)}
\newrgbcolor{userLineColour}{1 0.4 0.4}
\newrgbcolor{userFillColour}{1 0.4 0.4}
\rput{90}(20,50){\psellipse[linecolor=userLineColour,fillcolor=userFillColour,fillstyle=solid](0,0)(2,1.5)}
\newrgbcolor{userLineColour}{1 0.4 0.4}
\newrgbcolor{userFillColour}{1 0.4 0.4}
\rput{90}(20,70){\psellipse[linecolor=userLineColour,fillcolor=userFillColour,fillstyle=solid](0,0)(2,1.5)}
\newrgbcolor{userLineColour}{1 0.4 0.4}
\newrgbcolor{userFillColour}{1 0.4 0.4}
\rput{90}(20,10){\psellipse[linecolor=userLineColour,fillcolor=userFillColour,fillstyle=solid](0,0)(2,1.5)}
\rput(10,70){$\ha$}
\rput(10,50){$\hb$}
\rput(10,30){$\hc$}
\rput(10,10){$\hd$}
\newrgbcolor{userLineColour}{0.4 0 0.6}
\pspolygon[linewidth=0.55,linecolor=userLineColour](59.77,45.67)(56.5,50.06)(60.23,54.33)(63.5,49.94)
\newrgbcolor{userLineColour}{0.4 0 0.6}
\pspolygon[linewidth=0.55,linecolor=userLineColour](59.77,25.67)(56.5,30.06)(60.23,34.33)(63.5,29.94)
\psline[arrowscale=1 1.45,arrowsize=1.45 2,arrowlength=1.55]{->}(23,29)(55,29)
\psline[border=0.9,arrowscale=1 1.45,arrowsize=1.45 2,arrowlength=1.55]{->}(23,51)(55,51)
\psline[border=0.9,arrowscale=1 1.45,arrowsize=1.45 2,arrowlength=1.55]{->}(23,49)(55,32)
\psline[border=0.9,arrowscale=1 1.45,arrowsize=1.45 2,arrowlength=1.55]{->}(23,31)(56,48)
\psline[border=0.75,arrowscale=1 1.45,arrowsize=1.45 2,arrowlength=1.55]{->}(22.5,11.88)(56.88,45.62)
\psline[border=0.75,arrowscale=1 1.45,arrowsize=1.45 2,arrowlength=1.55]{->}(23.12,9.38)(55.62,26.25)
\rput[t](30,12){$\seven$}
\rput[b](27,34){$\four$}
\rput[t](27,46){$\three$}
\rput[b](91,43){$\nine$}
\rput[t](91.88,53.12){$\ten$}
\rput[t](91,36){$\eleven$}
\rput[t](90,20){$\twelve$}
\rput(60,50){$\Three$}
\rput(60,30){$\One$}
\rput[b](60,56.88){$\concone$}
\rput[t](60,23.12){$\conctwo$}
\end{pspicture}
\caption{Decomposition of $M$}
\label{fig-decomp}
\end{minipage}
\end{figure}
The new nodes in-between the shopping agents $\Sbj$ and the shops $\Obj$ emerge from the trust concept spectrum $\Ent_M = \{\Phi_1, \Phi_2\}$. The decomposition shows that,  e.g., $c$'s total trust rating 1.12 for the shop $i$ turns out to consist of a positive {\em qualified\/} trust rating $.5\times 3 \times .83 = 1.24$ for the concept $\Phi_1$ and a negative qualified component $.3\times 1\times (-.4) = -.12$ for the concept $\Phi_2$. The concepts induce the qualified trust matrices
\bear
F_1 & = &  \begin{pmatrix}
.83 \\
.55  \\
0  
\end{pmatrix} \begin{pmatrix} .5 & .5 & .5 & .5 \end{pmatrix} \ \ =\ \  \begin{pmatrix}  .41 & .41 & .41 & .41\\
.27 & .27 & .27 & .27\\
0&0&0&0 \end{pmatrix} \\
F_2 & = &  \begin{pmatrix}
-.4 \\
.6  \\
7  
\end{pmatrix} \begin{pmatrix} 0 & .5 & .3 & -.8 \end{pmatrix} \ =\ \begin{pmatrix}  0 & -.2 & -.12 & .32\\
0 & .3 & .18 & -.48\\
0&.35&.21&-.56 \end{pmatrix}
\eear
The similarity preserving matrix from Prop.~\ref{sim-decomp} is $F = F_1+F_2$. On the other hand, the trust matrix $M$ is decomposed as $M = 3F_1 + F_2$.

What has the banker learned by decomposing the trust matrix in this way? Whether he was aware of the trust concepts $\Phi_1$ and $\Phi_2$ or not, they are the intrinsic coordinate axes of trust in his town. It may be that the shoppers in the World Wild West mainly shop for two things, say guns and food (like in the Old Wild West). If the trust ratings in the matrix $M$ are proportional to the value of the transactions from which they are derived, then the trust concept $\Phi_1$, corresponding to the singular value $\lambda_1=3$, may correspond to guns, and the trust concept $\Phi_2$ to food, with the singular value $\lambda_2=1$, reflecting the fact that the average cost in a gun purchase is 3 times higher than the average cost in a food purchase. If the trust ratings are based on user feedback, then the eigenvalues tell that the shoppers assign 3 times higher value to guns; or that the gun community is 3 times more influential than the food community. Most shoppers and most shops belong to both communities, but the former one is more important for them. 

In any case, the banker now reads off from the matrix decomposition that shop $i$ sells high quality  guns, and carries the highest qualified trust rating .83 for $\Phi_1$; but that it also carries very bad food, as the negative qualified trust rating of -.4 for $\Phi_2$ shows. On the other hand, the shop $k$, whose total trust ratings are the lowest of all, actually supplies the best food, as reflected by the highest qualified trust rating .7 for $\Phi_2$. So if the banker wants to support an efficient market, he will recommend the shopper $b$ to go to $k$ if he needs to fill his virtual pantry, and to $i$ if he needs a gun. In general, the shoppers shopping for food should use the qualified trust matrix $F_1$; the shoppers shopping for guns should use the qualified trust matrix $F_2$.

\subsection{Effectiveness and scalability of the approach} 
The described spectral decomposition is based on the Singular Value Decomposition. The Singular Value Decomposition of a large matrix $M$ can be effectively computed,  up to any desired precision, by the standard iterative methods. Kleinberg's HITS algorithm \cite{kleinberg99authoritative} is an example of such a method, previously used for recognizing web spam \cite{TrustRank}. If a finite precision suffices, then a matrix can be decomposed much faster, in a single sweep, by bidiagonalization followed by some eigenvalue algorithm. Bidiagonalization of an $\Obj\times \Sbj$-matrix $M$ is quadratic in $\Sbj$ and linear in $\Obj$ \cite[Chap.~31]{Trefethen:NLA}. The eigenvalues of bidiagonal matrices are computed in linear time. Several versions of this approach, applicable to large examples, are implemented, e.g., in the GNU Scientific Library \cite[Sec.~14.4]{GNU}.

\section{Final comments and future work}\label{Conclusion}
\subsection{Summary}
In \cite{PavlovicD:FAST08} and in Sec.~\ref{Paradox} we saw that the fragile distribution of trust arises from the fact that it is like money: money makes money, and trust attracts more trust. In fact, the roles of money and of trust in a market essentially depend on their main shared feature, that they are both abstractions of past interactions. Both trust and money are acquired in exchange for concrete goods and services; but these origins are "laundered", abstracted away, so that money and trust can be used for any type of goods or services\footnote{Vespasian pointed out that {\em "Pecunia non olet"\/}: the money that he collected from the taxes on urinals bears no smell of its origin.}. In the case of money, this abstraction is what allows the flow of the investments towards more profitable areas. In the case of trust, this abstraction allows trust transfer, which is needed to facilitate new interactions. But it also leads to the vulnerabilities of trust, and to the paradox of the trust services, discussed in the Introduction. 

Arguably, the scale-free distribution of trust facilitates its adverse selection; but it is actually the abstraction of trust that enables this adverse selection. Moreover, the profitability of farming, selling and hijacking trust is increased because the trust ratings accumulated for one service can be used for another one. It is therefore not surprising that a market of trust is rapidly developing among the web merchants, e.g. on eBay and Amazon. The shops with high ratings and good feedback are often sold to new owners, for whom the accumulated trust has a higher value. They usually keep the same name and appearance, but they are consolidated into chains, and sometimes offer new goods and services. The difference in the valuation of trust between the old and the new owners is in some cases due to economy of scale: the chains are more profitable because they have lower maintenance costs.  
%In other cases, larger merchants may have a higher trust valuation because only trust accumulated above a threshold attracts customers.
In other cases a higher trust valuation arises from using the purchased trust for a more profitable type of service. Some trusted services are acquired by scammers, who have a very high valuation for trust. Hence adverse selection.

One way to make trust more reliable is thus to make it less abstract, by binding it to the trust concept for which it was accumulated. We showed how trust concepts can be reconstructed from the structure of a trust graph even when only records the scores. Such a trust network, recording the trust between some users and some providers\footnote{The users and the providers may or may not be drawn from the same set.} can be built collaboratively, {\em or\/} by surveillance of the network transactions, indexing network links, with or without a direct participation of the users. 

%We presented a method to decrease vulnerability of trust by mining trust networks for trust concepts, and by qualifying each trust rating as a linear combination of these basic concepts. The space of trust concepts a common subspace of the space of trustor communities on one side and the space of trustees communities on the other side. Each trust network uniquely determines such a subspace. Its Singular Value Decomposition yields basic trust concepts as eigenspaces, i.e. the subspaces of maximal correlation between the trustors and the trustees. The resulting qualified trust ratings can be used to guide the network interactions in a more precise way, which decreases the incentive for their transfer and abuse.

For simplicity, we presented the trust mining algorithm on a trust network spanned by the trust relations $a\truss{} r \ell$, with the trust concepts initially abstracted away. The trust concepts $\Phi$ were then reconstructed by mining. However, for the price of some technical complexity, the same approach extends to trust relations in the form $a\truss \Phi r \ell$, with the trust concepts previously qualified from a given lattice of concepts $\Ent$ given in advance. After a period of trust building, recorded in a trust network, the lattice $\Ent$ is refined to capture the additional trust concepts that emerge from the new interactions. In this way, the concept lattice $\Ent$ evolves.
%Instead of a matrix of trust ratings $a\truss{} r \ell$, this approach would mine a matrix of trust vectors $a\trusss{}{r_1, r_2,\ldots r_n}\ell $, which capture the trust relationships $a\truss{\Phi_i}{r_i} \ell$ for $\Phi_i=1,2,\ldots ,n$, like in Sec.~\ref{qualified}. By extending the described mining algorithm, we further decompose these trust vectors, and qualify trust in a finer lattice of trust concepts that arise from the recorded interactions. These finer trust  concepts are now obtained as subspaces of the spaces of tensors $\Rat^{\SSbj\times \Ent}$ and $\Rat^{\OObj\times \Ent}$.
While $\Ent$ can be thought of as a propositional logic of concept, it should be noted that is obtained as a lattice of subspaces of a vector space.  Such lattices are modular, but not distributive. In other words, the logic of concepts does not support a deduction theorem. This phenomenon is known to reflect the presence of hidden variables. Explaining it, and developing concept logic in general, is an interesting task for future work. But there is much more.

\subsection{Future work}
The problem of extrapolating the missing trust ratings, predicting how much Alice can trust Bob for her particular requirements, and providing a trust recommendation even if they never met, is probably the most important part of mining for trust. There is no space to continue with that task in the present paper. It should be noted, though, that the approach through similarity networks and spectral decomposition of their morphisms provides a crucial tool for this task. It arises from the universal property of the spectral decomposition. If the similarity morphisms $f_0:A_0\to B_0$ and $f_1:A_1\to B_1$ are
are both contained within a \emph{partial} similarity morphism $f:A\to B$, along the inclusions of $A_0$ and $A_1$ in $A$, and of $B_0$ and $B_1$ in $B$, then the universal property implies that the spectral decompositions of $f_0$ and $f_1$ can be \emph{glued together}. This allows us to extrapolate the values of $f$ from the available parts of its spectral decomposition. A similar gluing of Singular Value Decompositions of submatrices of a sparse matrix is hampered by their different singular values.

\bibliographystyle{plain}
\bibliography{ref-trust,../PavlovicD}

%\appendix
%\section*{Appendix: Proofs}
%\bprf{ of Lemma \ref{unique}.}
%
%\epr

\end{document}